\documentclass[aps,preprintnumbers,amsmath,amssymb,12pt,floatfix]{revtex4}
\usepackage{graphicx}
\usepackage{euscript}
\usepackage{float}
\usepackage{subfigure}

\begin{document}

\title{Theory of rigid-plane phonon modes in layered crystals}
\date{\today}
\author{K.H. Michel$^{1,*}$ and B. Verberck$^{1,2,\circ}$}
\affiliation{$^1$Departement Fysica, Universiteit Antwerpen, Groenenborgerlaan 171, 2020 Antwerpen, Belgium \\
$^2$Institut f\"{u}r Theoretische Physik und Astrophysik,
Universit\"{a}t W\"{u}rzburg, Am Hubland, D-97070 W\"{u}rzburg, Germany \\
$^*$E-mail: ktdm@skynet.be \\
$^\circ$E-mail: bart.verberck@ua.ac.be}

\begin{abstract}
The lattice dynamics of low-frequency rigid-plane modes in metallic (graphene multilayers, GML) and in insulating (hexagonal boron-nitride multilayers, BNML) layered crystals is investigated.  The frequencies of shearing and compression (stretching) modes depend on the layer number ${\EuScript N}$ and are presented in the form of fan diagrams.  The results for GML and BNML are very similar.  In both cases only the interactions (van der Waals and Coulomb) between nearest-neighbor planes are effective, while the interactions between more distant planes are screened.  A comparison with recent Raman scattering results on low-frequency shear modes in GML [Tan {\it et al.}, 
arXiv:1106.1146v1 (2011)] is made. Relations with the low-lying rigid-plane phonon dispersions in the bulk materials are established.  Master curves which connect the fan diagram frequencies for any given ${\EuScript N}$ are derived.  Static and dynamic thermal correlation functions for rigid-layer shear and compression modes are calculated.  The results might be of use for the interpretation of friction force experiments on multilayer crystals.
\end{abstract}

\maketitle

\section{Introduction}
The experimental discovery of graphene and other free-standing two-dimensional (2D) crystals \cite{Nov1,Nov} has opened the path for the synthesis of a whole class of layered materials with novel physical properties and with a great potential for technological applications \cite{GeiScience}.  The most prominent member, graphene --- a monoatomic layer of crystalline C with hexagonal structure --- is a metallic conductor.  In addition to an unusual electronic spectrum, this material shows extraordinary mechanical strength \cite{Lee} and thermal properties \cite{Balandin}.

On the other hand 2D hexagonal boron-nitride (h-BN) is an insulator \cite{Nov1,Nov} (3D h-BN has a direct band gap in the ultraviolet region \cite{Watanabe}).  While graphene is a purely covalent crystal, 2D h-BN, built from III-V elements, has partially covalent and ionic bonds.  The ionic character is a consequence of the charge transfer of $\approx 0.6$ electrons from B to N \cite{Schwarz}.  Since the crystal structure of 2D h-BN is non-centrosymmetric (point group symmetry $D_{3h}$), the two sublattices (B$^+$ and N$^-$) exhibit an electromechanical coupling.  Hence 2D h-BN is the structurally most simple crystal which, according to theoretical predictions \cite{MicPRB2009}, should be piezoelectric.

Nanoscale thin sheets of graphene, 2D h-BN and related layered materials \cite{Zabel} are of great importance for applications as electronic devices and nanoelectromechanical systems.  The synthesis and characterization of multilayers and the study of their physical and chemical properties is a challenge of current solid-state physics and materials science.  In particular, the change of properties with the number of layers and the evolution of the layer system to the corresponding bulk material are of foremost importance.  Most remarkable is for instance the change in electronic structure from graphene, a zero-gap semiconductor, to graphite, a semimetal with band overlap \cite{Par}.  These theoretical results are directly related to the interpretation of electronic transport experiments \cite{Nov1,Nov,Zhang}.  The change in the electronic bands is reflected in the double resonance Raman spectrum that clearly evolves with the number of layers \cite{Ferrari}.  Beside the electronic structure, the elastic properties depend on the number of layers.  Atomic force microscopy experiments (AFM) on various thin-sheet materials demonstrate that the nanoscale friction decreases with increasing number of layers \cite{Lee2}.  It has been suggested that the trend arises from the thinner sheets' increased susceptibility to out-of-plane elastic deformations.

Since physical properties vary with the number of layers, it is important to study separately the lattice dynamics of modes where the atomic planes move as rigid units.  Since this motion is governed by the weak interlayer forces, the corresponding frequencies are low ($\lesssim  150$ cm$^{-1}$) in comparison with the high-frequency optical modes ($\lesssim 1600$ cm$^{-1}$) which are due to covalent intralayer forces.  In graphite these low-frequency modes have been discovered half a century ago by inelastic neutron scattering experiments \cite{Dol}.  One distinguishes modes where the planes move parallel to the hexagonal axis (we will call these modes compression modes), and modes where the planes move perpendicular to this axis (shear modes).  Later on the complete phonon dispersions associated with rigid-plane motion have been measured by neutron scattering \cite{Nic}.  The rigid-plane (-layer) shear mode is optically active and has been measured by Raman scattering in graphite \cite{Nem2} and in 3D h-BN \cite{Nem3}.  Due to the weakness of the interlayer forces, the
rigid-layer shear frequency in graphite \cite{Han} and in h-BN \cite{Kuz} increases strongly with applied pressure.

The measurement of rigid-layer modes in few-layer systems has been an outstanding problem.  Neutron scattering is not an adequate technique since the samples are too small.  Most recently, the interlayer shear modes in few-layer graphene systems have been uncovered by Raman spectroscopy \cite{Tan}.  The increase of the resonance frequency with increasing layer number provides a unique signature for few-layer graphene systems and for multilayers in general.

In the present paper we report on theoretical studies of low-frequency rigid-layer shear modes and compression modes in graphene- and boron-nitride multilayers.  While the high-frequency optical mode spectra of graphene- and boron-nitride multilayers are very different due to the efficiency of Coulomb forces in the latter \cite{MicPRB2011}, the low-frequency optical mode spectra in both systems turn out to be very similar.

The content of the paper is as follows.  First (Sect.\ II) we present the main theoretical concepts which are used to treat by analytical means the lattice dynamics of multilayer systems.  Within a same formalism we consider graphene multilayers and h-BN multilayers.  The phonon dispersion relations of the corresponding rigid-layer motions (compression and shear modes) are calculated in Sect.\ III.  The dependence of the frequency spectra on the number of layers ${\EuScript N}$ is presented in the form of fan diagrams.  Next (Sect.\ IV) we compare the theoretical results with experiment.  Then we derive relations between the phonon frequencies of the rigid layer systems and the dispersions of the corresponding bulk materials.  We derive master curves which allow to connect the fan diagram frequencies for any given ${\EuScript N}$.  In Sect.\ V we calculate static and dynamic thermal displacement correlation functions.  The temperature dependence of displacement correlations of surface layers is calculated, the dependence on the layer number ${\EuScript N}$ is investigated.  Concluding remarks (Sect.\ VI) close the paper.

\section{Lattice dynamics}
\begin{figure}
\resizebox{4cm}{!}{\includegraphics{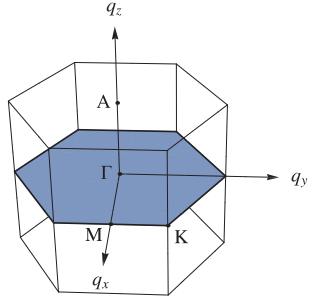}}
\caption{
Brillouin zone of the 3D hexagonal primitive lattice $\Gamma_h$; the shaded hexagon containing the $\Gamma$, M and K points is the Brillouin zone of a 2D hexagonal crystal.}
\label{fig1}
\end{figure}

In previous work we have studied by analytical methods the phonon dispersion relations for graphene multilayers (GML) \cite{MicPRB2008} and h-BN multilayers (BNML) \cite{MicPRB2011}.  Here we briefly recall the main concepts.  Both graphene and 2D h-BN have the same symmetry $D_{3h}$ with two atoms per unit cell.  Since each atom has three degrees of freedom, the dynamical matrix ${\EuScript D}(\vec{q}_\perp)$ for the planar problem has dimension $6\times 6$.  Here $\vec{q}_\perp$ is the wave vector in the 2D Brillouin zone (Fig.\ \ref{fig1}).  The 3D parent crystals, graphite and bulk h-BN have the same space group symmetry, $P6_3/mmc$ ($D_{6h}^4$).  Since in both cases there are 4 atoms per unit cell, the corresponding dynamical matrices are of dimension $12\times 12$.  In the case of GML, electron diffraction experiments \cite{Ferrari} have shown that the stacking of atomic planes is the same as in graphite ($\hdots$ABAB$\hdots$) \cite{Dre}.  In 3D h-BN, each B atom is on top of a N atom in the adjacent plane and vice versa, with $\hdots$AA$'$AA$'\hdots$ stacking \cite{Pea}.  We assume that the same holds for BNML.

We will use a unified description of the dynamical matrix for GML and BNML; the differences in structure and in interatomic (ionic) forces will be taken into account in the numerical evaluation of the secular equation.  We consider a slab of a finite number of ${\EuScript N}$ layers, the layers are labelled by an index $l\in\left\{0,1,\hdots,{\EuScript N} - 1 \right\}$.  The distance between nearest-neighbor planes which are perpendicular to the crystallographic $\vec{c}$ axis is $c/2$.  Since the slab is indefinitely extended only in two dimensions, we regard it as a 2D crystal which consists of prismatic unit cells \cite{All} with basis area $a^2 \sqrt{3}/2$ and height ${\EuScript N}c/2$.  We recall that $a = |\vec{a}_1| = |\vec{a}_2|$ is the length of the lattice translation vectors of the 2D hexagonal basis crystal \cite{Dre}.  Each unit cell contains ${\EuScript N}$ pairs of atoms (C,C) or (B,N) in the case of GML or BNML, respectively.  Since each atom has three vibrational degrees of freedom $i$ $(j)\in \{x,y,z \}$, the ${\EuScript N}$-layer slab has $6{\EuScript N}$ vibrational modes.

In order to calculate the phonon dispersion relations, we construct the $6{\EuScript N} \times 6{\EuScript N}$ dynamical matrix $\Delta_{\EuScript N}(\vec{q}_\perp)$.  In terms of $6\times 6$ submatrices ${\EuScript D(l,l'|\vec{q_\perp})}$ with elements ${\EuScript D}_{ij}^{\kappa \kappa'}(l,l'|\vec{q}_\perp)$, where $\kappa$ ($\kappa'$) takes two values $\kappa$ ($\kappa'$) $\in \{\text{C},\text{C} \}$ or $\{ \text{B}, \text{N}\}$ which corresponds to $\{\text{C},\text{C}\}$ or $\{\text{B}, \text{N}\}$, the dynamical matrix reads
\begin{align}
  \Delta_{\EuScript N}(\vec{q}_\perp) = \left(\begin{array}{ccccc}
  {\EuScript D}(0,0|\vec{q}_\perp) & {\EuScript D}(0,1|\vec{q}_\perp) & {\EuScript D}(0,2|\vec{q}_\perp) & \hdots & {\EuScript D}(0,{\EuScript N} - 1|\vec{q}_\perp)  \\
  {\EuScript D}(1,0|\vec{q}_\perp) & {\EuScript D}(1,1|\vec{q}_\perp) & {\EuScript D}(1,2|\vec{q}_\perp) & \hdots & {\EuScript D}(1,{\EuScript N} - 1|\vec{q}_\perp)  \\
  {\EuScript D}(2,0|\vec{q}_\perp) & {\EuScript D}(2,1|\vec{q}_\perp) &  {\EuScript D}(2,2|\vec{q}_\perp) & \hdots & {\EuScript D}(2,{\EuScript N} - 1|\vec{q}_\perp)  \\
					 \vdots & \vdots & \vdots & \ddots & \vdots \\
					 {\EuScript D}({\EuScript N} - 1,0|\vec{q}_\perp)  & {\EuScript D}({\EuScript N} - 1,1|\vec{q}_\perp)  & {\EuScript D}({\EuScript N} - 1,2|\vec{q}_\perp)  & \hdots & {\EuScript D}({\EuScript N} - 1,{\EuScript N} - 1|\vec{q}_\perp) 
					 \end{array}\right). \label{eq23}
\end{align}
Here we take into account the interaction within a same layer ($l = l'$) and interactions between layers separated by a distance $(l - l')c/2$, $l\ne l'$.  The ``same-plane'' matrices ${\EuScript D}(l,l|\vec{q}_\perp)$ are given by
\begin{align}
  {\EuScript D}(l,l|\vec{q}_\perp) = D(l,l|\vec{q}_\perp) + K(l,l|\vec{q}_\perp = 0),
\end{align}
where $D(l,l|\vec{q}_\perp)$ is the dynamical matrix of the $l$-th monolayer while $K(l,l|\vec{q}_\perp = \vec{0})$ accounts for the self-interaction due to interplane couplings.  Assuming that the in-plane interactions are the same for all planes, one has in terms of elements
\begin{align}
  D_{ij}^{\kappa \kappa'}(l,l|\vec{q}_\perp) = F_{ij}^{\kappa\kappa'}(l,l|\vec{q}_\perp) + C_{ij}^{\kappa \kappa'}(l,l|\vec{q}_\perp).
\end{align}
The matrices $F$ and $C$ stand for the in-plane covalent and Coulomb interactions, respectively.  In the case of GML, only covalent interactions are taken into account, $F\ne 0$, $C = 0$; in the case of BNML, both $F$ and $C$ are taken into account.  The interplane coupling matrices ${\EuScript D}(l,l'|\vec{q}_\perp)$, $l\ne l'$ in Eq.\ (\ref{eq23}) are due to van der Waals and Coulomb interactions $J$ and $C$, respectively:
\begin{align}
  {\EuScript D}_{ij}^{\kappa \kappa'}(l,l'|\vec{q}_\perp) = J_{ij}^{\kappa\kappa'}(\vec{q}_\perp)\delta_{l',l\pm1} + C_{ij}^{\kappa \kappa'}(l,l'|\vec{q}_\perp).
\end{align}
Here again Coulomb interactions are relevant for BNML with $1 \le |l - l'| \le {\EuScript N} - 1$, while for both BNML and GML van der Waals forces act between nearest neighbor planes only.  Hence in the case of GML, only nearest-neighbor off-diagonal elements ${\EuScript D}(l,l\pm 1|\vec{q}_\perp)$ are non-zero in Eq.\ (\ref{eq23}) (see Eq.\ (27) of Ref. \cite{MicPRB2008}). 

In calculating the submatrices $F(l,l|\vec{q}_\perp)$ we take into account intra-plane covalent interactions by means of a force-constant model originally derived from in-plane inelastic X-ray scattering experiments on single crystals of graphite \cite{Moh}.  In case of BNML, intra- and extra-plane Coulomb interaction matrices are calculated by means of Ewald's method \cite{Mar}.  We have solved numerically the secular determinant of order $6{\EuScript N}$,
\begin{align}
  |1\omega^2 - \Delta_{\EuScript N}(\vec{q}_\perp)| = 0, \label{deteq}
\end{align}
and obtained the phonon dispersion relations for GML \cite{MicPRB2008} and BNML \cite{MicPRB2011}.
In terms of eigenvalues $\bigl(\omega_\lambda(\vec{q}_\perp) \bigr)^2$ and orthogonal eigenvectors $\vec{\xi}_\lambda(\vec{q}_\perp)$ we have
\begin{align}
  \bigl(\omega_\lambda(\vec{q}_\perp) \bigr)^2 = \sum_{ll'}\sum_{\kappa\kappa'}\sum_{ij}\xi_i^{(l,\kappa)*}(\lambda,\vec{q}_\perp)
   {\EuScript D}_{ij}^{\kappa\kappa'}(l,l'|\vec{q}_\perp)\xi_j^{(l',\kappa')}(\lambda,\vec{q}_\perp).
\end{align}
Here $\lambda$ labels the $6{\EuScript N}$ eigenmodes.  We discern 3 acoustic modes such that $\omega_\lambda(\vec{q}_\perp = \vec{0}) = 0$ for $\lambda = 1,2,3$ and $6{\EuScript N} - 3$ optical modes with $\omega_\lambda(\vec{q}_\perp) \ne 0$ for all values of $\vec{q}_\perp$.  Among the latter we distinguish near $\vec{q}_\perp = \vec{0}$ $3{\EuScript N}$ atomic vibrational modes (in-plane and out-of-plane displacements) and $3({\EuScript N} - 1)$ rigid-plane modes.

As has been shown previously \cite{MicPRB2011}, marked differences between GML and BNML appear in the highest optical branches with frequencies $\approx 1300$ -- $1500$ cm$^{-1}$.  These modes are due to intra-plane shear displacements where the Coulomb forces in BNML are efficient.  On the other hand for a given ${\EuScript N}$, the low-frequency ($<200$ cm$^{-1}$) optical phonon dispersions in GML and BNML are very similar \cite{MicPRB2011,MicPRB2008}.  These modes are due to rigid-plane compression and shear displacements.  Due to the overall charge neutrality of the BN atomic planes, the Coulomb forces between nearest-neighbor planes are screened and hence there is no qualitative difference between GML and BNML rigid-plane modes.  In the following section we will discuss the evolution of the rigid-layer modes with the number of layers.

\section{Rigid-layer modes}
We first recall the situation in the 3D parent materials graphite and h-BN.  Since the point group symmetry is $D_{6h}$, the decomposition into irreducible representations of the optical displacements at the center of the Brillouin zone (Fig.\ \ref{fig1}) reads \cite{Man,Nem} $\Gamma = A_{2u} + 2B_{2g} + E_{1u} + 2E_{2g}$.  While six of these modes are high-frequency inter-plane vibrational modes ($800$ -- $1600$ cm$^{-1}$), one of the doubly degenerate $E_{2g}$ modes and one $B_{2g}$ mode refer to low-frequency
rigid-plane motions.  We denote these modes by $E_{2g_1}$ and $B_{2g_1}$.  The $E_{2g_1}$ mode corresponds to the rigid-plane shear displacements perpendicular to the crystallographic $\vec{c}$ axis.  This mode has been measured in graphite by Raman reflectivity \cite{Nem} at $42 \pm 1$ cm$^{-1}$ and is called ``rigid-layer shear'' mode.  It can be identified with the zero wave vector transverse optical mode (TO) near $1.35$ THz measured first by neutron scattering in high-quality pyrolytic graphite \cite{Nic}.  The $B_{2g_1}$ mode corresponds to a rigid-layer compression mode along the $\vec{c}$ axis.  This mode is optically inactive, however it appears near $3.9$ THz ($\approx 130$ cm$^{-1}$) in neutron scattering \cite{Nic}.  For a more complete discussion of the early work, we refer to Ref. \cite{Nem2}.  A discussion of the zone-center optical modes in h-BN was originally given in Ref.\ \cite{Geick}, the $E_{2g_1}$ rigid-layer shear mode was observed by Raman scattering \cite{Nem3} at $51.8$ cm$^{-1}$. Recently the phonon dispersions of h-BN have been measured by inelastic x-ray scattering and analyzed by ab initio calculations \cite{SerBos}.  At the $\Gamma$ point, the $E_{2g_1}$ rigid-layer shear mode has energy $6.5$ meV ($52$ cm$^{-1}$), the compression rigid-layer mode (called there $B_{1g}$) has energy $15$ meV ($121$ cm$^{-1}$).

In the ${\EuScript N}$-layer system the $3({\EuScript N} - 1)$ rigid-plane optical modes at the center of the 2D Brillouin zone decompose into ${\EuScript N} - 1$ compression modes with frequencies $\omega_{\lambda_c}(\vec{q}_\perp = \vec{0})$, $\lambda_c = 1, 2, 3, \hdots, {\EuScript N} - 1$ and eigenvectors $\vec{\xi}(\lambda_c,\vec{q}_\perp = \vec{0})$ and into ${\EuScript N} - 1$ doubly degenerate shear modes with frequencies $\omega_{\lambda_s}(\vec{q}_\perp = \vec{0})$ and eigenvectors $\vec{\xi}(\lambda_s^{(1)},\vec{q}_\perp = \vec{0})$ or $\vec{\xi}(\lambda_s^{(2)},\vec{q}_\perp = \vec{0})$, where $\lambda_s^{(1)}$ ($\lambda_s^{(2)}$) $= 1, 2, 3, \hdots, {\EuScript N} - 1$.
The degeneracy of the shear modes is a consequence of hexagonal symmetry at $\vec{q}_\perp = \vec{0}$.  The compression and shear modes correspond respectively to the low-frequency modes $B_{2g_1}$ and $E_{2g_1}$ of the bulk materials.  Since the $\lambda_c$ modes have only nonzero displacement components along the $z$-direction ($\vec{c}$-axis) while the $\lambda_s$ modes have only nonzero $x$- and $y$-components, one has the simplified orthonormality conditions
\begin{align}
  \sum_{l,\kappa} \xi_z^{(l,\kappa)}(\lambda_c,\vec{0}) \xi_{z}^{(l,\kappa)}(\lambda'_c,\vec{0}) & = \delta_{\lambda_c \lambda'_c}, \label{eq7} \\
  \sum_{i}\sum_{l,\kappa} \xi_i^{(l,\kappa)}(\lambda_s,\vec{0}) \xi_{i}^{(l,\kappa)}(\lambda'_s,\vec{0}) & = \delta_{\lambda_s \lambda'_s}, \label{eq8}
\end{align}
where $i\in\{x,y\}$.  The eigenvector components fulfill the relations
\begin{align}
\frac{\xi_i^{(l,\kappa)}(\lambda_\alpha,\vec{0})}{\sqrt{m_\kappa}}
 = \frac{\xi_i^{(l,\kappa')}(\lambda_\alpha,\vec{0})}{\sqrt{m_{\kappa'}}},\text{ } \kappa\ne\kappa', \label{eq9}
\end{align}
where $m_\kappa$ is the mass of particle $\kappa$, $\lambda_\alpha = \lambda_c$ ($\lambda_s$) for $i = z$ ($x,y$).  Hence particles within a same plane $l$ experience equal displacements.  In addition one has
\begin{align}
\sum_l \xi_i^{(l,\kappa)}(\lambda_\alpha,\vec{0}) = 0, \label{eq10}
\end{align}
where $\alpha = s$ for $i\in\{x,y \}$ and $\alpha = c$ for $i = z$, which means that the center of mass of the ${\EuScript N}$-layer system stays at rest.

\begin{figure}
\resizebox{8cm}{!}{\includegraphics{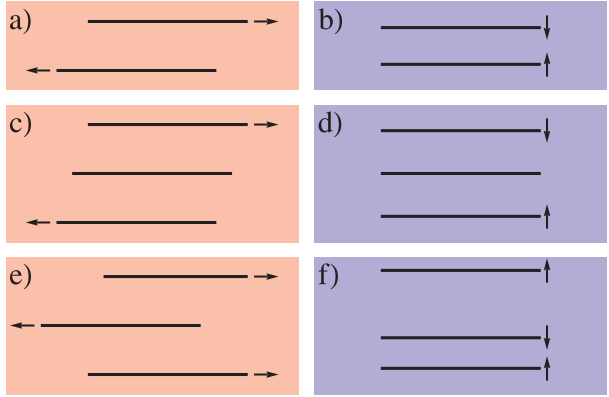}}
\caption{
Rigid-plane shear and compression displacements, ${\EuScript N} = 2$, (a) and (b), respectively; ${\EuScript N} = 3$, (c), (e) and (d), (f), respectively.
\label{fig2}}
\end{figure}

As an example we first consider GML.  For the case
${\EuScript N} = 2$ (bilayer), we obtain the rigid-plane shear mode eigenvectors with 12 components
\begin{align}
  \vec{\xi}_{\EuScript N = 2}(x) & = \frac{1}{2}(1,0,0,1,0,0,-1,0,0,-1,0,0), \\
  \vec{\xi}_{\EuScript N = 2}(y) & = \frac{1}{2}(0,1,0,0,1,0,0,-1,0,0,-1,0),
\end{align}
with degenerate eigenfrequency $30.4$ cm$^{-1}$ [see Fig.\ \ref{fig2}(a)].  Here the
components $1 - 3$ and $4 - 6$ refer to the Cartesian displacements of the first ($\kappa = 1$) and second ($\kappa = 2$) particle in plane $l = 1$, respectively, while the components $7 - 9$ and $10 - 12$ refer to the displacements of the two particles in the plane $l = 2$.
The rigid-plane compression mode eigenvector for the bilayer reads
\begin{align}
  \vec{\xi}_{\EuScript N = 2}(z) = \frac{1}{2}(0,0,1,0,0,1,0,0,-1,0,0,-1),
\end{align}
with eigenfrequency $90.1$ cm$^{-1}$ [see Fig.\ \ref{fig2}(b)].

\begin{figure}
\resizebox{8.5cm}{!}{\includegraphics{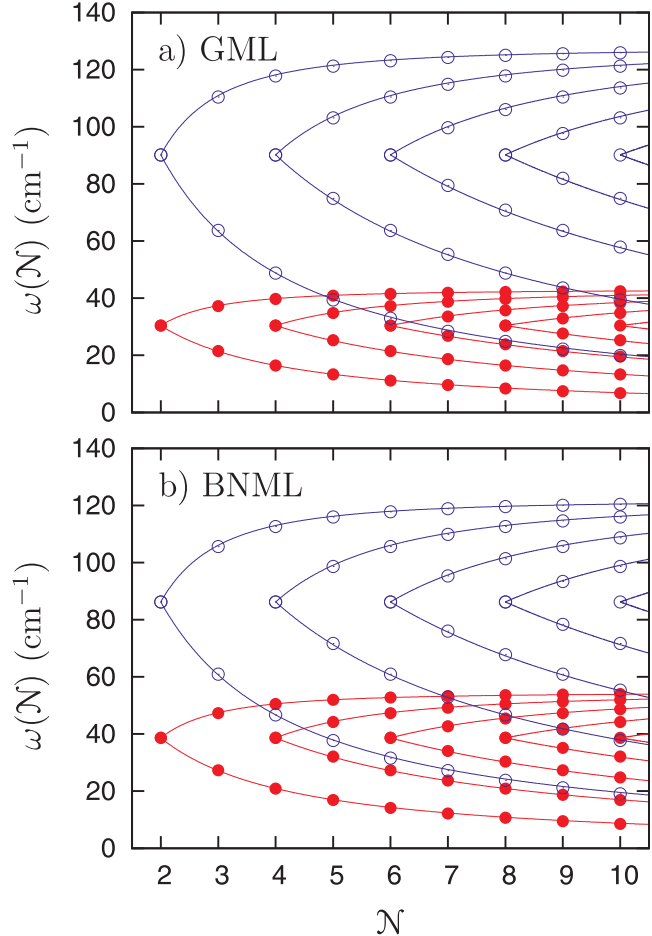}}
\caption{Rigid-layer frequencies $\omega({\EuScript N})$ at the $\Gamma$ point for (a) GML and (b) BNML as a function of number of layers ${\EuScript N}$.
Lines connecting the frequency points are master curves given by Eqs.\ (\ref{mastera}) and (\ref{masterb}).
For both GML and BNML, the two sets of points (red, filled circles and blue, open circles) are referred to as lower and upper fan diagrams, respectively.  
\label{fig3}}
\end{figure}

For the case ${\EuScript N} = 3$  there are two doubly degenerate shear modes with frequencies $21.5$ cm$^{-1}$ and $37.2$ cm$^{-1}$, shown respectively in Figs.\ \ref{fig2}(c) and (e), and two compression modes with frequencies $63.7$ cm$^{-1}$ and $110.4$ cm$^{-1}$, Figs.\ \ref{fig2}(d) and (f).  We notice that for one degenerate shear mode and one compression mode [Figs.\ \ref{fig2}(c), (d)] the center plane undergoes no displacement.  This feature is characteristic for all multilayers with ${\EuScript N}$ uneven.  On the other hand the displacements of the two outer layers are opposite to the displacements of the inner (central) layer [Figs.\ \ref{fig2}(e) and (f)], in agreement with the general requirement that the center of mass of the ${\EuScript N}$-layer system stays at rest.  We have plotted the calculated frequencies of the rigid-plane shear modes and compression modes for GML as a function of the layer number ${\EuScript N}$ in the form of fan diagrams in Fig.\ \ref{fig3}(a).  The lower set (red, filled circles), centered around the bilayer shear frequency $30.4$ cm$^{-1}$, corresponds to the rigid-layer shear modes $\{\omega_{\lambda_s} ({\EuScript N})\}$; the upper set (blue, open circles), centered around the bilayer compression frequency $90.1$ cm$^{-1}$ corresponds to the rigid-layer compression modes $\{\omega_{\lambda_c} ({\EuScript N}) \}$.  In the fan diagram of shear motions for GML the sequence of frequencies at $30.4$ cm$^{-1}$ which occurs for even values of ${\EuScript N}$ refers to the situation where the $\frac{{\EuScript N}}{2}$ upper planes of the system move in unison in a direction perpendicular to the hexagonal axis ($\vec{c}$), while the $\frac{{\EuScript N}}{2}$ lower planes move in unison in opposite direction.  The same holds for the compression motions where the frequencies at $90.1$ cm$^{-1}$ in the fan diagram correspond to two sets of unison motions with opposite direction along $\vec{c}$.

Proceeding along the same lines, we have calculated the rigid-layer eigenfrequencies and displacement vectors for the case of BNML.  Shown in Fig.\ \ref{fig3}(b) are again the eigenfrequency fan diagrams as a function of ${\EuScript N}$.  The lower set (red, filled circles) corresponds to the rigid-layer shear modes and is centered around the BN bilayer shear frequency $38.6$ cm$^{-1}$, the upper set (blue, open circles), centered around the BN bilayer compression frequency $86.3$ cm$^{-1}$ corresponds to the rigid-layer compression modes.

In the next section we discuss these results and compare with recent experiments and ab initio calculations on GML \cite{Tan}.

\section{Discussion of results}
\subsection{Comparison with experiment}
Given the similarity of the fan diagrams (Figs.\ \ref{fig3}(a) and (b) for GML and BNML, respectively) we first discuss some general features.  We recall that for the GML systems we have only taken into account van der Waals forces between nearest-neighbor planes.  On the other hand we have treated the BNML systems as ionic crystals.  In addition to van der Waals forces between nearest-neighbor planes we have summed the Coulomb interactions over all planes.  Obviously the overall charge neutrality and the layer rigidity, i.e. the absence of relative motion between the B$^+$ and N$^-$ sublattices, leads to a screening of the Coulomb interactions between next-nearest-neighbor and more distant h-BN planes.  Hence the low-frequency fan diagrams for GML and BNML are very similar.  We recall that the screening effect disappears for non-rigid-layer displacements, as is seen from the high-frequency ($\ge 1300$ cm$^{-1}$) optical mode dispersions in BNML \cite{MicPRB2011} which are qualitatively very different from the corresponding dispersions in GML \cite{MicPRB2008}.

\begin{figure*}
\resizebox{14cm}{!}{\includegraphics{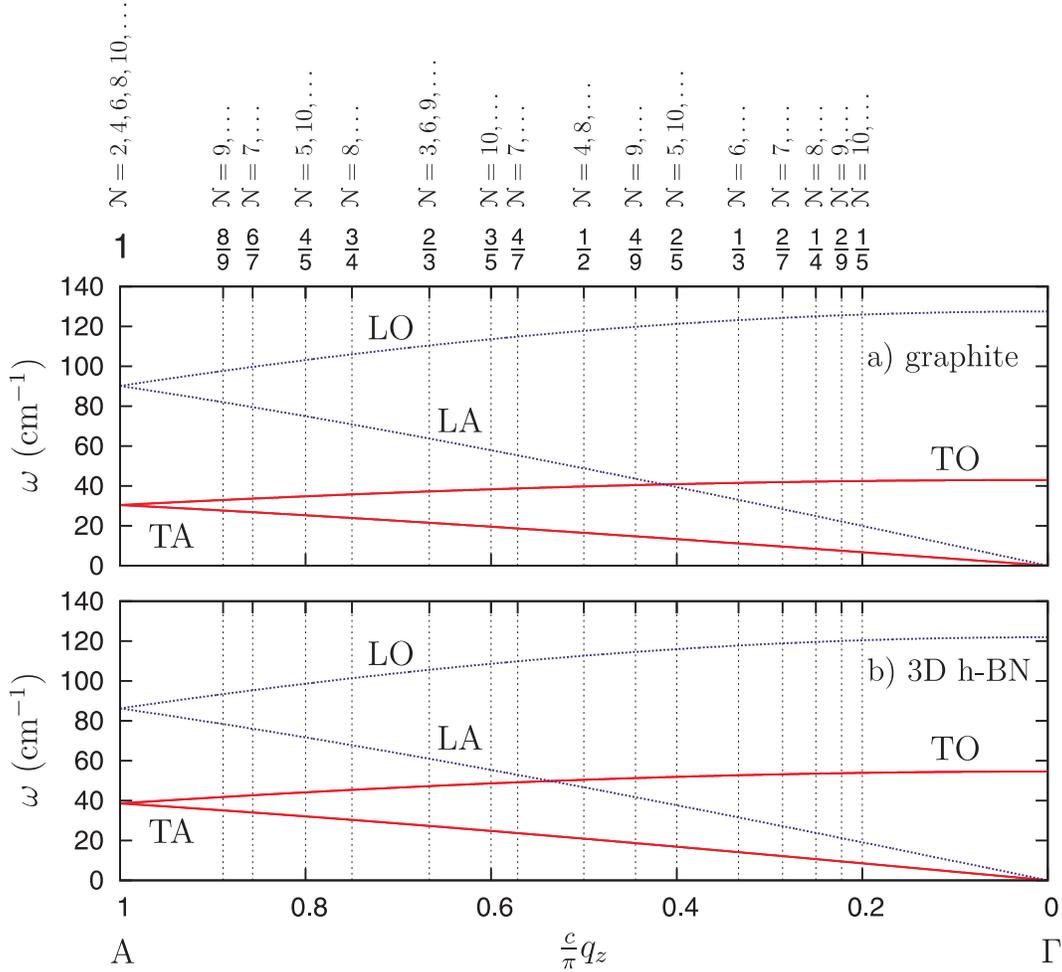}}
\caption{Low-frequency phonon branches along A--$\Gamma$ ($\vec{q} = (0,0,q_z):(0,0,\frac{\pi}{c})\longrightarrow \vec{0}$) for (a) graphite and (b) 3D h-BN.  Cuts at well-defined values of $q_z$ (marked by vertical lines and labels at the top horizontal axis) yield the $\Gamma$ point frequencies of the multilayers (see text).
\label{fig4}}
\end{figure*}

The similarity of the low-frequency results of GML and BNML is also reflected by the low-frequency spectra of the 3D parent materials.  
We recall that transverse acoustic (TA) and optical (TO) as well as longitudinal acoustic (LA) and optical (LO) phonon dispersions along $\vec{q} = (0,0,q_z)$ were first measured by neutron scattering \cite{Nic} in pyrolytic graphite and recently by inelastic X-ray scattering in single crystals of graphite \cite{Moh} and in h-BN \cite{SerBos}.
In Figs.\ \ref{fig4}(a) and (b) we have plotted the low-frequency phonon dispersion relations for graphite \cite{MicPRB2008} and for bulk h-BN \cite{MicPRB2011} along the line A--$\Gamma$ in the 3D Brillouin zone (see Fig.\ \ref{fig1}).  Similar results have been obtained earlier by first-principles calculations for graphite \cite{Wir,Pav} and h-BN \cite{Ker}.

We first discuss the low-frequency dispersions of graphite in relation with the GML low-frequency shear modes.  The two lowest branches (TA and TO) in Fig.\ \ref{fig4}(a) (red, full lines) refer to the acoustic and optical rigid-plane shear motion, respectively.  The optical branch TO evolves from the frequency value $30.4$ cm$^{-1}$ at the A point of the Brillouin zone to $43.0$ cm$^{-1}$ at the $\Gamma$ point.  We notice that the value at the A point [$\vec{q}_\text{A} = (0,0,\frac{\pi}{c})$] agrees with the rigid-plane shear frequency $30.4$ cm$^{-1}$ of the bilayer at $\vec{q}_\perp = \vec{0}$ and with the center points for even ${\EuScript N}$ in the lower fan diagram of Fig.\ \ref{fig3}(a) (red, filled circles).
We attribute this agreement to the fact that in our calculation on graphite and multilayer systems we have restricted ourselves to van der Waals interactions between nearest-neighbor planes only \cite{MicPRB2008}.  However, since this is a well-justified approximation, this correspondence should also be valid experimentally (inelastic neutron scattering on graphite \cite{Nic} gives $\nu \lesssim 1$ THz at the A point, most recent low-frequency Raman experiments \cite{Tan} measure a shear frequency of $31$ cm$^{-1}$ for bilayer graphene).
The value $43.0$ cm$^{-1}$ of the $E_{2g_1}$ mode at the $\Gamma$ point of the 3D system corresponds to the limit value for large ${\EuScript N}$ of the sequence of highest frequencies $\{\omega_{\lambda_s}^h(\EuScript N) \} =
\{30.4$ cm$^{-1}$; $37.2$ cm$^{-1}$; $39.7$ cm$^{-1}$; $\hdots$; $42.4$ cm$^{-1}; \hdots\}$ for ${\EuScript N} = \{ 2; 3;
4; \hdots ; 10; \hdots \}$, respectively.

In case of h-BN [Fig.\ \ref{fig4}(b)] the rigid-layer shear mode TO evolves from $38.6$ cm$^{-1}$ at A to $54.5$ cm$^{-1}$ at $\Gamma$ (symmetry $E_{2g_1}$).
Notice that the value at A agrees again with the shear mode frequency of the bilayer.
Here the Coulomb interaction is only effective between nearest-neighbor layers and screened between more distant layers in the bulk system and in ${\EuScript N}$-layer systems.
The value at $\Gamma$ agrees with the corresponding limit frequencies $\{\omega_{\lambda_s}^h(\EuScript N) \}$ of BNML for ${\EuScript N}\longrightarrow \infty$.

The sequence of frequencies $\{\omega_{\lambda_s}^h(\EuScript N) \}$ has been measured recently in GML up to ${\EuScript N} = 11$ by polarized Raman techniques \cite{Tan}.  These experiments demonstrate that the shift of the resonance (called C peak \cite{Tan}) with ${\EuScript N}$ is truly representative of the GML system.  Furthermore the authors of Ref.\ \cite{Tan} have studied the eigenfrequencies and eigenvectors of the ${\EuScript N}$-layer shear modes by using a simple linear-chain model and by performing ab initio calculations up to ${\EuScript N} = 5$.  Comparing our own calculated values of the shear-mode eigenfrequencies with those of Ref.\ \cite{Tan}, we see close agreement.

It is useful to trace back this agreement on the level of interlayer van der Waals (vdW) interactions.  From experiment the authors of Ref.\ \onlinecite{Tan} derive that the interlayer force constant per unit area $\alpha$ has the value $\alpha \sim 12.8\times 10^{18}$ N/m$^3$.  The area of the unit cell in graphene is $v_\text{2D} = a^2\sqrt{3}/2 = 5.24\times 10^{-16}$ cm$^2$ ($a = 2.46$ {\AA}).  We then define the interlayer force constant per unit cell $\tilde{\alpha} = \alpha v_\text{2D}$ with value $670$ dyn/cm.  In Ref.\ \onlinecite{Tan} the rigid-layer shear frequency of graphite is obtained as the limit for ${\EuScript N}\longrightarrow \infty$ of the GML frequency $\omega_{\EuScript N}$ and reads $\omega_\infty = 2\sqrt{\alpha/\mu}$.  Here $\mu = 2M/v_\text{2D} = 7.6\times 10^{-8}$ g/cm$^2$ is the mass per unit area ($M = 12$ u for C).  Writing then $\omega_\infty = \sqrt{2\tilde{\alpha}/M}$, we compare with the expression of the rigid-layer shear mode in graphite $\omega(E_{2g_1})$, Eq.\ (31) of Ref.\ \onlinecite{MicPRB2008}.   It is straightforward to recast this expression in the form $\omega(E_{2g_1}) = \sqrt{2\tilde{h}_{xx}/M}$, where $\tilde{h}_{xx}$ is a sum of interatomic vdW force constants between next-neighbor graphene planes.  With the values of Ref.\ \cite{MicPRB2008} we obtain $\tilde{h}_{xx} = 654$ dyn/cm, comparable with 670 dyn/cm \cite{Tan}.  For the graphene bilayer we obtain $\omega_{\lambda_s}({\EuScript N} = 2) = \sqrt{\tilde{h}_{xx}/M}$ and hence $\omega(E_{2g_1}) = \sqrt{2} \omega_{\lambda_s}({\EuScript N} = 2)$.  We recall that the values of these interlayer shear force constants have to be chosen ad hoc, they can not be obtained from currently accepted Lennard-Jones potentials for C--C vdW interactions \cite{Gir}.  Similar relations hold for the rigid-layer compression modes.  We obtain 
$\omega(B_{2g_1}) = \sqrt{2 \tilde{h}_{zz}/M}$ and for the bilayer $\omega_{\lambda_c}({\EuScript N} = 2) = \omega (B_{2g_1})\sqrt{2}$.  The force constant is given by $\tilde{h}_{zz} = 5755$ dyn/cm.

\subsection{Relation with bulk dispersions}
We now establish quantitative relations between the shear-mode eigenfrequencies $\{\omega_{\lambda_s}(\EuScript N) \}$ of the
multilayers GML and BNML
and the TA and TO dispersions with 
wave vector $\vec{q} = (0,0,q_z)$ along A--$\Gamma$ in graphite and h-BN, respectively.  With the ${\EuScript N}$-layer system we associate a quantized wave length $\nu \lambda_z = {\EuScript N}c/2$ or equivalently a wave vector $q_z^{\EuScript N}(\nu) = 4\pi \nu /{\EuScript N} c$, where $\nu$ is an integer in the interval $1 \le \nu \le {\EuScript N}/2$.
The center of mass of the multilayer system stays at rest in displacements associated with the wave vector $\vec{q}_z^{\EuScript N}(\nu)$.  We calculate the sum of phase factors of the corresponding displacement pattern,
\begin{align}
  \sum_{n = 0}^{{\EuScript N} - 1} e^{iq_z^{\EuScript N}(\nu) \frac{n}{2}c} = \frac{\sin(\pi \nu)e^{i({\EuScript N - 1)}\frac{\pi \nu}{\EuScript N}}}{\sin(\frac{\pi \nu}{\EuScript N})},
\end{align}
which is zero since $\nu$ is an integer in the interval $[1,{\EuScript N}/2]$.
Since the nearest distance between equivalent planes in the bulk materials is $2(c/2)$, we have the correspondence
\begin{align}
  q_z(\nu) \equiv \frac{q_z^{\EuScript N}(\nu)}{2} = \frac{2\pi}{c}\frac{\nu}{\EuScript N}. \label{relqz}
\end{align} 
We find that the ${\EuScript N} - 1$ rigid-layer shear modes $\{\omega_{\lambda_s}({\EuScript N})\}$ of the ${\EuScript N}$
layer-system are obtained as the intersections of the two lowest phonon branches TA and TO in Fig.\ \ref{fig4}(a) (red, full lines) with vertical lines located at $\vec{q} = \bigl(0,0,q_z(\nu)\bigr)$, $\nu$ integer $\in [ 1, \frac{\EuScript N}{2} ]$.  For ${\EuScript N}$ even, the value $\nu = \frac{{\EuScript N}}{2}$ corresponds to $\vec{q} = (0,0,\frac{\pi}{c})$, i.e.\ the A point of the 3D Brillouin zone.  Hence one obtains the series of central points with constant frequencies
in the fan diagrams.  These are located at $\omega = 30.4$ cm$^{-1}$
for the shear
modes in GML
[see Fig.\ \ref{fig3}(a)], and at $\omega = 38.5$ cm$^{-1}$
for the
shear modes in BNML [Fig.\ \ref{fig3}(b)].
The cuts of the graphite A--$\Gamma$ branches resulting in the GML fan diagrams for ${\EuScript N} \le 10$ are shown in Fig.\ \ref{fig4}(a) by vertical lines at $q_z$ given by Eq.\ (\ref{relqz}); the values of $\frac{c}{\pi}q_z = \frac{2\nu}{{\EuScript N}}$
as well as the corresponding ${\EuScript N}$-values
are labeled on the top horizontal axis.  Note that the frequencies present for ${\EuScript N}$ are also present for all multiples of ${\EuScript N}$.  Also note that in the limit ${\EuScript N} \longrightarrow \infty$ a cut of the graphite
A--$\Gamma$ branches at $q_z = 0$, i.e.\ the $\Gamma$ point, is reached, consistent with the observation that the
frequency of the $E_{2g_1}$ mode at the $\Gamma$ point ($\omega = 43.0$ cm$^{-1}$) corresponds to the limit value of the highest frequencies $\{ \omega_{\lambda_s}^h ({\EuScript N})  \}$ of the lower fan diagram of Fig.\ \ref{fig3}(a) (red, filled circles) mentioned before.
On the other hand,
the lowest frequencies $\{ \omega_{\lambda_s}^l ({\EuScript N})  \}$ of the lower fan diagram of Fig.\ \ref{fig3}(a) (red, filled circles) evolve to zero for ${\EuScript N} \longrightarrow \infty$, in agreement with the limit of the TA branch with $q_z \longrightarrow 0$. 
We next consider the rigid-plane compression modes $\{\omega_{\lambda_c}({\EuScript N})\}$  in GML, upper fan diagram of Fig.\ \ref{fig3}(a) (blue, open circles).  Here too, the relation [Eq.\ (\ref{relqz})] with the LA and LO modes in graphite can be established.  The two branches meet at the A point at $90.1$ cm$^{-1}$ which agrees with the value of the bilayer and rigid-plane compression eigenfrequency at $\vec{q}_\perp = \vec{0}$.  The frequency $\omega(B_{2g_1}) = 127.5$ cm$^{-1}$ of the LO branch at $\Gamma$ in graphite corresponds to the limit for large ${\EuScript N}$ of the sequence of highest eigenfrequencies of the compression modes in GML
$\{\omega_{\lambda_c}^h ({\EuScript N})\} = \{90.1$ cm$^{-1}$; $110.4$ cm$^{-1}$; $117.8$ cm$^{-1}$; $\hdots$; $125.9$ cm$^{-1}; \hdots \}$ for
${\EuScript N} = \{ 2 ; 3 ; 4; \hdots;  10; \hdots \}$, respectively.
Here the lowest frequencies $\{ \omega_{\lambda_c}^l ({\EuScript N}) \}$ tend to zero for ${\EuScript N} \longrightarrow \infty$, in agreement with the TA branch.
  The evolution of this mode with increasing number of layers has already been studied in one of our previous papers \cite{MicPRB2008}.  
Since the $B_{2g_1}$ mode in graphite is optically silent, this prediction has not been checked by experiment.  However, from a group-theoretical analysis it has been concluded \cite{Sah} that in GML with ${\EuScript N}$ even there are ${\EuScript N}/2$ compression modes with symmetry $A_{1g}$ that are Raman active.  Also for the case of ${\EuScript N}$ uneven, infrared active modes should occur.
Again, the eigenfrequencies plotted in the upper fan diagram in Fig.\ \ref{fig3}(a) coincide with the intersections of the vertical lines $\vec{q} = (0,0,q_z)$ with $q_z$ given by Eq.\ (\ref{relqz}) with the LO and LA phonon branches along A--$\Gamma$ in graphite.
From a comparison of low-frequency out-of-plane phonon dispersions of ${\EuScript N}$-layer graphene at $\Gamma$ and the low-frequency dispersions LO, LA of graphite along $\Gamma$--A, it has been inferred \cite{Kar} that a relation like Eq.\ (\ref{relqz}) holds, however with $\nu = 0, 1, \hdots, {\EuScript N} - 1$.  This range of $\nu$ overestimates the number of rigid plane compression modes by more than a factor 2.

We now turn to BNML.  Comparing with the low-frequency dispersions TA, TO and LA, LO of h-BN along A--$\Gamma$ in the Brillouin zone, we find that the intersections obtained by means of Eq.\ (\ref{relqz}) determine again the eigenfrequencies of the multilayers.  This holds as well for the shear modes $\{\omega_{\lambda_s}({\EuScript N})\}$ as for the compression modes $\{\omega_{\lambda_c} ({\EuScript N})\}$ (lower and upper fan diagrams in Fig.\ \ref{fig3}(b), respectively)  and is illustrated in Fig.\ \ref{fig4}(b).

Having established the relation between the A--$\Gamma$ phonon branches TO, TA and LO, LA of the 3D materials and the eigenfrequencies $\{\omega_{\lambda_s}({\EuScript N})\}$ and $\{\omega_{\lambda_c}({\EuScript N}) \}$ at the $\Gamma$ point of the multilayers
for rigid-shear and rigid-compression modes
[Eq.\ (\ref{relqz})], it is possible to deduce $\omega({\EuScript N})$ master curves connecting the fan diagram frequencies (Fig.\ \ref{fig3}).
The following considerations hold for the shear modes as well as for the compression modes.  In the former case $\omega_\text{A}$ and $\omega_\Gamma$ stand for $\omega_\text{A}^\text{TO}$ and $\omega_\Gamma^\text{TO}$ respectively, in the latter case for
$\omega_\text{A}^\text{LO}$ and $\omega_\Gamma^\text{LO}$.
First, one needs to consider each fan diagram as a set of pairs of $\omega({\EuScript N})$ curves; at $({\EuScript N},\omega) = (2n,\omega_\text{A})$, with $n = 1, 2, 3, \hdots$, a pair of curves (one increasing, the other decreasing) originates. 
From Eq.\ (\ref{relqz}) it follows that the frequencies lying on the curves originating at $(\omega_\text{A},2n)$ are obtained by cutting the corresponding A--$\Gamma$ branches of the 3D material at $Q_n \equiv \frac{c}{\pi}q_z(n) = \frac{2 n}{\EuScript N}$, with ${\EuScript N} = 2, 3, 4, \hdots$.  The curves containing the highest and lowest frequencies ($n = 1$) are e.g.\ obtained by cuts at $Q_1 = \frac{2}{2}, \frac{2}{3}, \frac{2}{4}, \hdots$.  Secondly, the A--$\Gamma$ 3D phonon branches can be extremely well approximated by second-degree curves.  For the optical branches, satisfying  $\omega(Q \equiv \frac{c}{\pi}q_z = 0) = \omega_\Gamma$, $\omega(Q = 1) = \omega_\text{A}$ and $\left. \frac{d\omega}{d Q}\right|_{Q = 0} = 0$, we put
\begin{align}
  \omega^+(Q) = \omega_\Gamma -(\omega_\Gamma - \omega_\text{A})Q^2.
\end{align}
For the acoustic branches, for which $\omega(Q = 0) = 0$ and $\omega(Q = 1) = \omega_\text{A}$, we assume
\begin{align}
  \omega^-(Q) = \omega_\text{A} Q \bigl[(1 + r) - rQ\bigr], \label{eq17}
\end{align}
with $r$ a dimensionless fit parameter.  The superscripts `$+$' and `$-$' refer to the increasing and the decreasing curves, respectively.  Now inserting $Q_n = \frac{2 n}{\EuScript N}$ results in
\begin{align}
  \omega^+_n({\EuScript N}) & = \omega_\Gamma -(\omega_\Gamma - \omega_\text{A})\frac{4n^2}{{\EuScript N}^2}, \label{mastera} \\
  \omega^-_n(\EuScript{N}) & = \omega_\text{A} \frac{2 n}{\EuScript N} \left[(1 + r) - r \frac{2 n}{\EuScript N}\right]. \label{masterb}
\end{align}
The $\omega^+_n({\EuScript N})$ and $\omega^-_n({\EuScript N})$ curves plotted in Figs.\ \ref{fig3}(a) and \ref{fig3}(b) have been obtained by evaluating Eqs.\ (\ref{mastera}) and (\ref{masterb}).  For the latter, values of $r = 0.165$ and $r = 0.166$ were fitted for GML and BNML, respectively.  The agreement between the multilayer phonon frequencies, obtained by diagonalizing the $6{\EuScript N} \times 6{\EuScript N}$ dynamical matrix, and the master curves, obtained by making cuts of the A--$\Gamma$ 3D phonon branches, is perfect.
Quadratic-form assumptions for the A--$\Gamma$ rigid-shear and rigid-compression phonon branches obviously work extremely well (for both graphite and h-BN).
Expression (\ref{eq17}) for $\omega^-(Q)$ is readily used to calculate the sound velocity of LA and TA phonons in the bulk materials.  Substituting $Q = \frac{c}{\pi} q_z$ in Eq.\ (\ref{masterb}) we obtain
\begin{align}
  V = \lim_{q_z \longrightarrow 0} \frac{\partial \omega^-}{\partial q_z} = \omega_\text{A} \frac{c}{\pi}(1 + r).
\end{align}
With $c = 6.7$ {\AA}, $\omega_\text{A}^\text{LA} = 90.1$ cm$^{-1}$ and $\omega_\text{A}^\text{TA} = 30.4$ cm$^{-1}$ we obtain in case of graphite the longitudinal and transverse sound velocities $V_\text{LA} = 4.22$ km/s and $V_\text{TA} = 1.42$ km/s.  The experimental values are $4.14(4)$ km/s and $1.48(6)$ km/s, respectively \cite{BosKri}.  For h-BN, with $c = 6.66$ {\AA}, 
$\omega_\text{A}^\text{LA} = 86.3$ cm$^{-1}$ and $\omega_\text{A}^\text{TA} = 38.6$ cm$^{-1}$, we obtain
$V_\text{LA} = 4.02$ km/s and $V_\text{TA} = 1.79$ km/s, to be compared with the experimental values \cite{BosSer} $3.44(3)$ km/s and $1.84(6)$ km/s, respectively.
Note further that the evolution of the rigid modes' highest frequencies (shear and compression) are given by \begin{align}
\omega^+_1({\EuScript N}) = \omega_\Gamma -(\omega_\Gamma - \omega_\text{A})\frac{4}{{\EuScript N}^2}.
\end{align}

The procedure of obtaining the phonon eigenfrequencies of the multilayer system by making intersections of the A--$\Gamma$ phonon branches of the 3D material is the analog of the zone-folding scheme where the phonon dispersion relations of 1D carbon nanotubes are obtained from 2D graphene \cite{Saibook,Reibook}.

\section{Displacement correlations}
Having determined the eigenfrequencies and eigenmodes for GML and BNML systems, we will calculate the temperature-dependent dynamic and static correlation functions.  The knowledge of these functions is relevant for the interpretation of scattering experiments and is likely to be useful for friction experiments \cite{Lee2}.  We use a quantum-mechanical formulation of lattice dynamics in the harmonic approximation and extend the standard theory of 3D crystals \cite{Mar} to the case of multilayers.  For a given ${\EuScript N}$-layer system we consider the time-dependent displacement operator $u_i(\vec{n}_\perp,l,\kappa;t)$.  Here $i\in\{ x,y,z\}$, $\vec{n}_\perp$ refers to the prismatic unit cell \cite{MicPRB2011}, $l$ to the multilayer plane, $0\le l \le {\EuScript N} - 1$, $\kappa$ to the particle of mass $m_\kappa$, $t$ stands for time.  The expansion in terms of normal coordinates $Q_\lambda(\vec{q}_\perp;t)$ reads
\begin{align}
u_i(\vec{n}_\perp,l,\kappa;t) = \frac{1}{\sqrt{N_\perp m_\kappa}}\sum_{\vec{q}_\perp,\lambda}\xi_i^{(l,\kappa)}(\lambda,\vec{q}_\perp)Q_\lambda(\vec{q}_\perp;t)e^{i \vec{q}_\perp \cdot \vec{X}(\vec{n}_\perp,l,\kappa)}.
\end{align}
Here $\vec{X}(\vec{n}_\perp,l,\kappa)$ is the equilibrium position of the particle $(\vec{n}_\perp,l,\kappa)$ in the multilayer crystal, $N_\perp$ the number of unit cells.  In terms of phonon creation $(b^\dagger)$ and annihilation $(b)$ operators the time-dependent normal coordinate reads
\begin{align}
  Q_\lambda(\vec{q}_\lambda;t) = \sqrt{\frac{\hbar}{2\omega_\lambda(\vec{q}_\perp)}}\left(b_\lambda^\dagger(-\vec{q}_\lambda)e^{i\omega_\lambda(\vec{q}_\perp)t} +  b_\lambda(\vec{q}_\lambda)e^{-i\omega_\lambda(\vec{q}_\perp)t}\right).
\end{align}
One has the usual commutation relations for Bose operators
\begin{align}
  \bigl[b_\lambda(\vec{q}_\perp),b_{\lambda'}^\dagger(\vec{q}_\perp')\bigr] & = \delta_{\vec{q}\vec{q_\perp'}}\delta_{\lambda\lambda'}, \\
  \bigl[b_\lambda(\vec{q}_\perp),b_{\lambda'}(\vec{q}_\perp)\bigr] & = \bigl[b_\lambda^\dagger(\vec{q}_\perp),b_{\lambda'}^\dagger(\vec{q}_\perp')\bigr] =0.
\end{align}
The thermal occupation of phonons with polarization $\lambda$ and frequency $\omega_\lambda(\vec{q}_\perp)$ is given by
\begin{align}
 \bigl\langle b_\lambda^\dagger(\vec{q}_\perp) b_{\lambda}(\vec{q}_\perp)  \bigr\rangle \equiv n_\lambda(\vec{q}_\perp) = \frac{1}{e^{\beta \hbar \omega_\lambda(\vec{q}_\perp) - 1}}, \label{eq25}
\end{align}
where $\beta = k_\text{B}T$, $T$ is the temperature and $k_\text{B}$ the Boltzmann constant.

In the case of rigid-layer displacements we retain at $\vec{q}_\perp = \vec{0}$ those eigenmodes that satisfy Eqs.\ (\ref{eq7}) -- (\ref{eq10}) and we denote the eigenfrequencies $\omega_{\lambda_\alpha}(\vec{q}_\perp = \vec{0})$ by $\omega_{\lambda_\alpha}$ and the occupation number $n_{\lambda_\alpha}(\vec{q}_\perp = \vec{0})$ by $n_{\lambda_\alpha}$.
Here again $\lambda_\alpha = \lambda_s$ refers to rigid-layer shear modes and $\lambda_\alpha = \lambda_c$ to rigid-layer compression modes. 
The rigid-layer displacement-displacement dynamical correlation function reads
\begin{align}
  \bigl\langle u_i(\vec{n}_\perp,l,\kappa;t)u_i(\vec{n}_\perp,l,\kappa;0)  \bigr\rangle_{\EuScript N} = \frac{1}{N_\perp m_\kappa} \sum_{\lambda_\alpha}\xi_i^{(l,\kappa)}(\lambda_\alpha,\vec{0}) \xi_i^{(l,\kappa)}(\lambda_\alpha,\vec{0})\bigl\langle Q_{\lambda_\alpha}(\vec{0};t) Q_{\lambda_\alpha}(\vec{0};0) \bigr\rangle_{\EuScript N}. \label{eq6}
\end{align}
We have taken into account that only terms diagonal in $\lambda_\alpha$ contribute to the thermal average $\langle \text{ }\rangle$.  Evaluation of the thermal average gives
\begin{align}
  \bigl\langle Q_{\lambda_\alpha}(\vec{0};t) Q_{\lambda_\alpha}(\vec{0};0) \bigr\rangle_{\EuScript N} = \frac{\hbar}{2\omega_{\lambda_\alpha}}\left[ n_{\lambda_\alpha} e^{i\omega_{\lambda_\alpha}t}  + (1 + n_{\lambda_\alpha})e^{-i\omega_{\lambda_\alpha}t}\right]. \label{eq27}
\end{align}
Notice that the right-hand side of Eq.\ (\ref{eq6}) is independent of $\vec{n}_\perp$ (rigid layers!).
In order to obtain the shear and compression correlation functions of the ${\EuScript N}$-layer system, we multiply both members of Eq.\ (\ref{eq6}) by the number of unit cells $N_\perp$, sum over $l$, $\kappa$ and $i \in \{ x,y \}$ or $i = z$.  We call the result $\bigl\langle u_\alpha(t) u_\alpha(0) \bigr\rangle_{\EuScript N}$ where $\alpha = s$ in case $i\in \{x,y\}$ and $\alpha = c$ for $i = z$.  The result reads
\begin{align}
  \bigl\langle u_\alpha(t) u_\alpha(0) \bigr\rangle_{\EuScript N} = f\sum_{\lambda_\alpha}\bigl\langle Q_{\lambda_\alpha}(t) Q_{\lambda_\alpha}(0) \bigr\rangle_{\EuScript N}, \label{eq28}
\end{align}
where
\begin{align}
  f = \sum_{l\kappa i} \frac{\xi_i^{(l,\kappa)}(\lambda_s,\vec{0}) \xi_i^{(l,\kappa)}(\lambda_s,\vec{0})}{m_\kappa}
    =  \sum_{l\kappa} \frac{\xi_z^{(l,\kappa)}(\lambda_c,\vec{0}) \xi_z^{(l,\kappa)}(\lambda_c,\vec{0})}{m_\kappa}
    = \frac{2}{m},  \label{res29}
\end{align}
with $m = \sum_\kappa m_\kappa$, $\kappa\in\{1,2 \}$.  The result (\ref{res29}) is a consequence of Eqs.\ (\ref{eq7}), (\ref{eq8}) and (\ref{eq9}).  Note that the result is independent of the number of layers; one has $f = 0.0806$ u$^{-1}$ or $0.0833$ u$^{-1}$ for BNML or GNML, respectively (with $m$ in atomic mass units u).  The sum over $\lambda_\alpha$ on the right-hand side of Eq.\ (\ref{eq28}) depends on ${\EuScript N}$.

\begin{figure}
\resizebox{8cm}{!}{\includegraphics{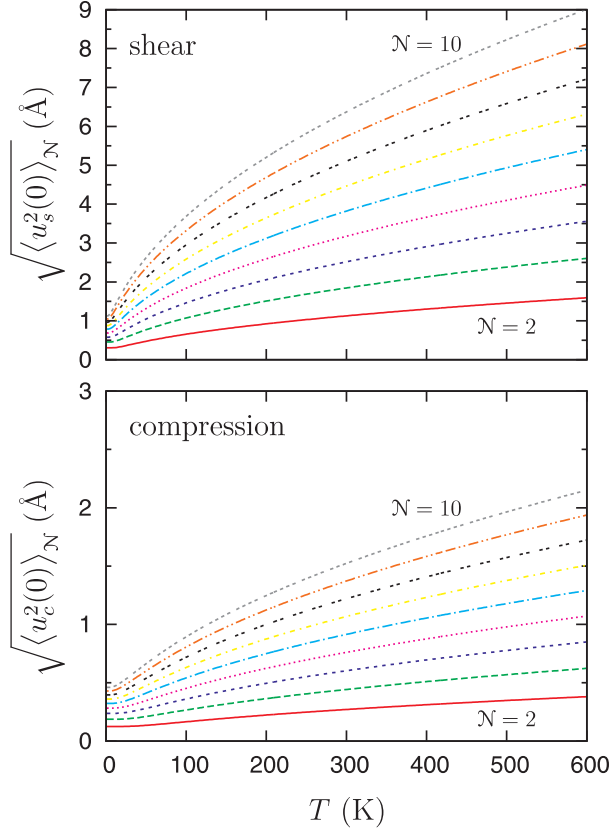}}
\caption{Temperature-dependent mean-square displacements $\sqrt{\bigl\langle u^2_\alpha(0) \bigr\rangle_{\EuScript N}}$ for rigid-plane shear ($\alpha = s$, top) and compression ($\alpha = c$, bottom) for GML.  The number of layers ${\EuScript N}$ ranges from $2$ (lowest curve) to $10$ (upper curve).  The results for BNML are very similar.
\label{fig5}}
\end{figure}

The static correlation functions are obtained by taking $t = 0$.  From Eq.\ (\ref{eq27}) we get, by using Eq.\ (\ref{eq25}),
\begin{align}
  \Bigl\langle \bigl(Q_{\lambda_\alpha}(\vec{0};0\bigr)^2  \Bigr\rangle_{\EuScript N} = \frac{\hbar }{2\omega_{\lambda_\alpha}} \coth \frac{\beta \hbar \omega_{\lambda_\alpha}}{2},
\end{align}
and hence
\begin{align}
  \bigl\langle u^2_\alpha(0) \bigr\rangle_{\EuScript N} = f \sum_{\lambda_\alpha} \frac{\hbar }{2\omega_{\lambda_\alpha}} \coth \frac{\beta \hbar \omega_{\lambda_\alpha}}{2}.
\end{align}
Making use of relations (\ref{mastera}) and (\ref{masterb}), we can write
\begin{align}
  \bigl\langle u^2_\alpha(0) \bigr\rangle_{\EuScript N} = gf\left\{ \frac{\hbar}{2\omega_\text{A}} \coth  \frac{\beta \hbar \omega_\text{A}}{2} + 
   \sum_{n} \left( \frac{\hbar}{2\omega_n^+({\EuScript N})} \coth  \frac{\beta \hbar \omega_n^+({\EuScript N})}{2}
    + \frac{\hbar}{2\omega_n^-({\EuScript N})} \coth  \frac{\beta \hbar \omega_n^-({\EuScript N})}{2} \right)\right\} \label{eq32}
    \end{align}
for ${\EuScript N}$ even ($n = 1, 2, 3, \hdots,\frac{{\EuScript N}}{2} - 1$) and
    \begin{align}
  \bigl\langle u^2_\alpha(0) \bigr\rangle_{\EuScript N} =  gf 
   \sum_{n} \left( \frac{\hbar}{2\omega_n^+({\EuScript N})} \coth  \frac{\beta \hbar \omega_n^+({\EuScript N})}{2}
    + \frac{\hbar}{2\omega_n^-({\EuScript N})} \coth  \frac{\beta \hbar \omega_n^-({\EuScript N})}{2} \right) \label{eq33}
\end{align}
for ${\EuScript N}$ odd ($n = 1, 2, 3, \hdots, \frac{{\EuScript N} - 1}{2}$).
Here $g = 2$ accounts for the degeneracy in case of the shear modes while $g = 1$ in case of compression modes.  We recall that Eqs.\ (\ref{eq32}) and (\ref{eq33}) apply to shear modes $\{ \lambda_s \}$ for $\omega_\text{A}$ and $\omega_\Gamma$ entering $\omega_n^{\pm}$ given by $\omega_\text{A}^\text{TO}$ and $\omega_\Gamma^\text{TO}$ and to compression modes $\{ \lambda_c \}$ for $\omega_\text{A}^\text{LO}$ and $\omega_\Gamma^\text{LO}$.  We have calculated the temperature-dependent mean-square displacements $\sqrt{\bigl\langle u^2_\alpha(0) \bigr\rangle_{\EuScript N}}$ for rigid-plane shear ($\alpha = s$) and compression ($\alpha = c$) modes by means of Eqs.\ (\ref{eq32}) and (\ref{eq33}) for GML and BNML.  Results for a series of ${\EuScript N}$-layer systems are shown in Fig.\ \ref{fig5}.  We recall that these results are obtained for rigid layers [Eq.\ (\ref{eq9})], while the center of mass of the ${\EuScript N}$-layer system stays at rest.  Hence the static displacement correlation functions $\bigl\langle u_s^2(0) \bigr\rangle_{\EuScript N}$ and
$\bigl\langle u_c^2(0) \bigr\rangle_{\EuScript N}$ are a measure of the total amount of the relative shear and compression motion, respectively, between rigid layers.

\begin{figure}
\resizebox{8cm}{!}{\includegraphics{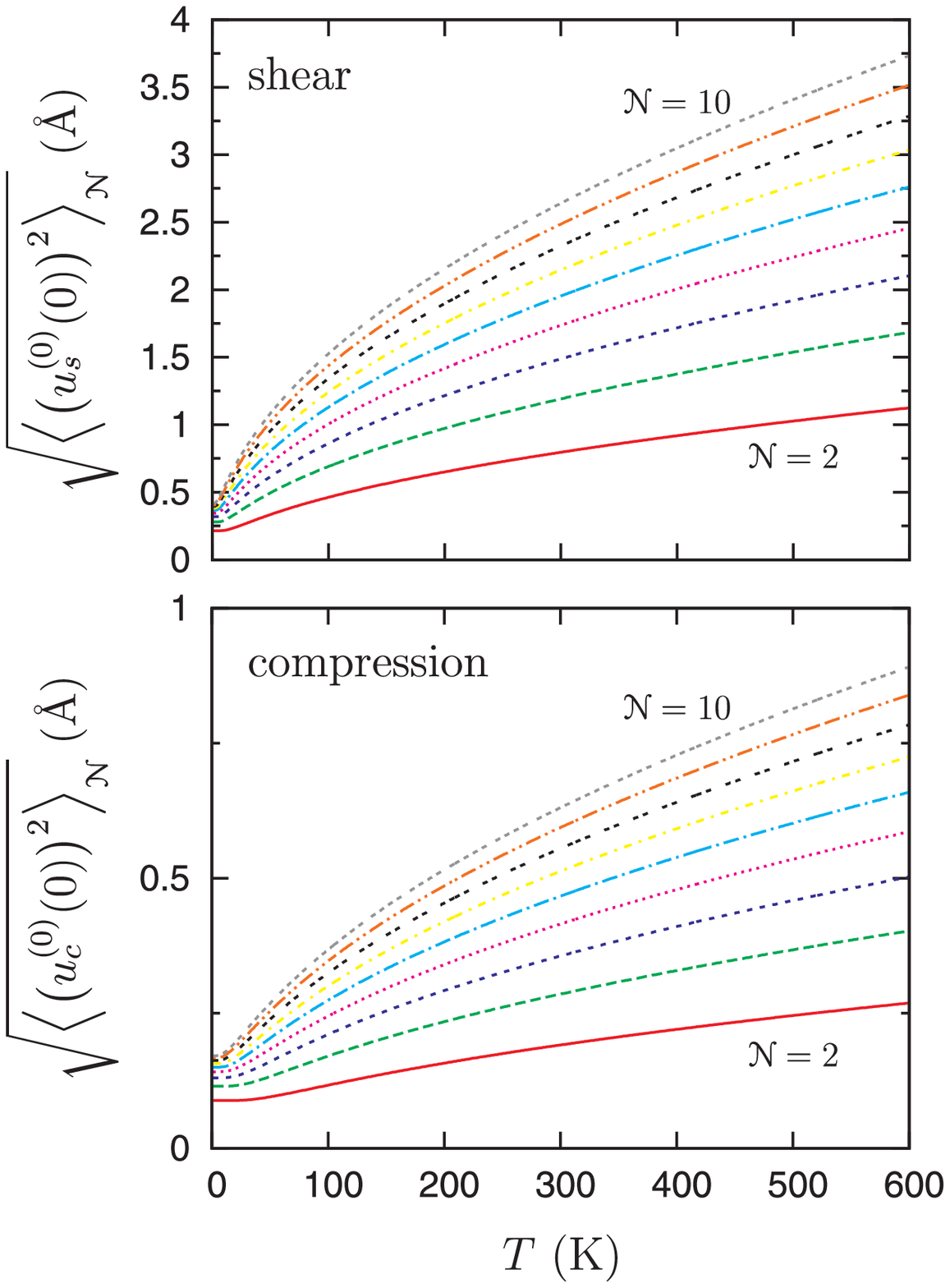}}
\caption{Temperature-dependent mean-square displacements $\sqrt{ \Bigl\langle \bigl(u^{(0)}_\alpha(0) \bigr)^2 \Bigr\rangle_{\EuScript N}}$ for rigid-plane shear ($\alpha = s$, top) and compression ($\alpha = c$, bottom) for GML.  The number of layers ${\EuScript N}$ ranges from $2$ (lowest curve) to $10$ (upper curve).  The results for BNML are very similar.
\label{fig6}}
\end{figure}

We next study the static displacement correlation function of the surface layer with label $l = 0$.  Instead of
Eq.\ (\ref{res29}), we have to consider
\begin{align}
  f_{\EuScript N}^{(0)} (\lambda_\alpha) = \sum_{i \kappa}\frac{\xi_i^{(0,\kappa)}(\lambda_\alpha,\vec{0}) \xi_i^{(0,\kappa)} (\lambda_\alpha,\vec{0})}{m_\kappa},
\end{align}
where $\alpha = s$ for $i \in \{ x, y\}$ and $\alpha = c$ for $i = z$.
The static correlation function now reads
\begin{align}
  \Bigl\langle \bigl(u^{(0)}_\alpha(0) \bigr)^2 \Bigr\rangle_{\EuScript N} = \sum_{\lambda_\alpha} 
    f_{\EuScript N}^{(0)} (\lambda_\alpha) \frac{\hbar }{2\omega_{\lambda_\alpha}} \coth \frac{\beta \hbar \omega_{\lambda_\alpha}}{2}.
\end{align}
In Fig.\ \ref{fig6} we show numerical results of the mean-square thermal displacements.

From Figs.\ \ref{fig5} and \ref{fig6} we conclude that the average rigid-layer shear and compression displacements increase with increasing temperature and with layer number ${\EuScript N}$.  In Ref.\ \onlinecite{Lee2} the results of friction force microscopy experiments demonstrate that friction decreases monotonically with the number of layers.  The mechanical origin for the observed effect is attributed to the fact that the sliding AFM tip causes out-of-phase deformations (puckering) of the surface sheet.  The increased tip-sheet contact area or (and) the additional work required to move the puckered region forward lead to increased friction.  This effect is more pronounced for thinner samples which exhibit a lower bending stiffness.  On the other hand for thicker sheets the puckering is less prominent owing to the larger bending stiffness of the sheet \cite{Lee2}.  Within this scenario it is suggested that some relative sliding between the topmost layer and the material below occurs.  This feature should increase the spacing of the stick-slip events.  Our results (Fig.\ \ref{fig6}) on the increase of the mean-square shear displacements of the surface layer with increasing ${\EuScript N}$ are then compatible with the experimental findings that the spacing of the stick-slip events increases with increasing ${\EuScript N}$ \cite{Lee2}.  Concerning the increase of the vertical (compression) mean-square displacements with increasing ${\EuScript N}$ we are led to argue that those processes decrease the contact area between AFM tip and multilayer system and hence contribute to a decrease of friction with increasing ${\EuScript N}$.

We close with a comment on dynamics.  The Fourier transform of the time-dependent correlation function $\bigl\langle u_\alpha(t) u_\alpha(0) \bigr\rangle$, $\alpha = s$ ($c$) is relevant for the interpretation of dynamic scattering laws.  We define
\begin{align}
  C_{{\EuScript N} {\EuScript N}}^{\alpha \alpha}(\omega) = \frac{1}{2\pi} \int_{-\infty}^{+\infty} dt \ e^{i \omega t} \bigl\langle u_\alpha(t) u_\alpha(0) \bigr\rangle_{\EuScript N}
\end{align}
and obtain by means of Eqs.\ (\ref{eq28}) and (\ref{eq27})
\begin{align}
  C_{{\EuScript N} {\EuScript N}}^{\alpha \alpha}(\omega) = f  \sum_{\lambda_\alpha} \frac{\hbar}{2 \omega_{\lambda_\alpha}}\left[ 
  n_{\lambda_\alpha} \delta(\omega + \omega_{\lambda_\alpha}) + (1 + n_{\lambda_\alpha}) \delta(\omega - \omega_{\lambda_\alpha})
  \right]. \label{expr35}
\end{align}
Here $\hbar \omega$ stands for the energy transfer of the scattering particle (photon or neutron) to the ${\EuScript N}$-layer system.  The first term within the square bracket represents an energy absorption by the scattering particle (anti-Stokes process) and the second term an energy loss (Stokes process) which becomes dominant at low $T$.  Expression (\ref{expr35}) comprises all shear or compression motion resonances of a given ${\EuScript N}$-layer system.  So far the highest value shear resonances $\{\omega_{\lambda_s}^h \}$ have been detected by experiment \cite{Tan}.  While these experiments have been carried out at room temperature, it might be necessary to go to lower $T$ in order to detect the resonances at lower frequencies.  Also the Raman-active compression modes in even ${\EuScript N}$ multilayers \cite{Sah}, symmetry $A_{1g}$, are a challenge for further experiments.

\section{Concluding remarks}
We have given a theoretical investigation of the low-frequency phonon dispersions in crystalline layered materials.  These phonons, 
associated with rigid-plane motions, show universal behavior which applies to metallic (GML) as well as to ionic, insulating (BNML) 
systems.  The frequency spectra have been represented in the form of fan diagrams for compression (also called stretching) and shearing 
motions.  For a system of ${\EuScript N}$ layers one distinguishes ${\EuScript N} - 1$ compression modes and ${\EuScript N} - 1$ doubly 
degenerate shear modes with frequencies $\{ \omega_{\lambda_c} ({\EuScript N})\}$ and $\{ \omega_{\lambda_s} ({\EuScript N})\}$, 
respectively.  The fan diagrams (see Fig.\ \ref{fig3}) are centered around a series of frequency points given by the bilayer frequencies $
\omega_{\lambda_c}({\EuScript N} = 2)$ and $\omega_{\lambda_s}({\EuScript N} = 2)$ appearing for systems with an even number of 
layers ${\EuScript N}$.  The fan diagram associated with compression is centered around higher frequencies than the fan diagram 
associated with shear motion in both GML and BNML.  For both shearing and compression the sequences of highest frequencies
$\{ \omega_{\lambda_s}^h ({\EuScript N}) \}$
and $\{ \omega_{\lambda_c}^h ({\EuScript N}) \}$ have as limits for ${\EuScript N} 
\longrightarrow \infty$ the bulk material frequencies $\omega(E_{2g_1})$ and $\omega(B_{2g_1})$ respectively at the $\Gamma$ point of 
the 3D Brillouin zone.  In case of GML the series of shear frequencies $\{ \omega_{\lambda_s}^h({\EuScript N})\}$ up to ${\EuScript N} = 
11$ has been measured by Raman scattering \cite{Tan}.  On the other hand the sequences of lowest frequencies $\{ \omega_{\lambda_s}^l ({\EuScript N}) \}$ 
and $\{ \omega_{\lambda_c}^l ({\EuScript N}) \}$ have limit values $0$ for ${\EuScript N} \longrightarrow \infty$.  Comparison with the
low-frequency dispersions along the $\Gamma$--A line in the Brillouin zone of hexagonal layered 3D materials shows that the frequencies 
$\omega_{\lambda_s}({\EuScript N} = 2)$ and $\omega_{\lambda_c}({\EuScript N} = 2)$ of the bilayer agree with the bulk frequencies $\omega$(TO) and $\omega$(LO), respectively, at the A point.  In addition one has the relations $\sqrt{2}\omega_{\lambda_c}({\EuScript N} = 2) = \omega(B_{2g_1})$ and $\sqrt{2}\omega_{\lambda_s}({\EuScript N} = 2) = \omega(E_{2g_1})$.  These relations are a consequence of the fact that the interlayer force constants $\tilde{h}_{zz}$ for compression and $\tilde{h}_{xx}$ for shearing are only effective between next-neighbor rigid planes.  Interactions between more distant rigid planes are negligible.  Note that for the case of rigid-layer
shear modes in GML this conclusion was drawn from Raman scattering results \cite{Tan}.  We have attributed the absence of longer distance interactions to screening effects.  In GML the screening is due to the metal nature ($\pi$ electrons), in BNML the screening is due to the overall charge neutrality and the plane rigidity.

We further have explored the relations between the LO and LA phonons of the bulk materials along $\vec{q} = [0,0,q_z]$ and the fan diagram frequencies for compression modes in ${\EuScript N}$-layer systems; similar relations exist between TO and TA phonons along A--$\Gamma$ and the fan diagram frequencies for shearing modes.  In both cases the ${\EuScript N} - 1$ rigid-layer frequencies $\{ \omega_{\lambda_c} ({\EuScript N})\}$ and $\{ \omega_{\lambda_s} ({\EuScript N}) \}$ are obtained as intersections of the LO, TA and TO, TA phonon branches, respectively, with vertical lines at discrete positions $q_z(\nu)$ along $\Gamma$--A.  We have obtained master curves which allow to derive the fan diagrams of GML and BNML for any given ${\EuScript N}$.

Finally we have calculated static and dynamic correlation functions for rigid-plane motions.  We have studied correlations as functions of $T$ and of ${\EuScript N}$.  Our results, which exhibit again a large similarity between GML and BNML, might be of relevance for the understanding on an atomistic level of the results of force friction experiments on thin-layer sheets \cite{Lee2}.

\acknowledgments
The authors are indebted to J. Maultzsch for bringing Ref.\ \onlinecite{Tan} to their attention.  They thank D. Lamoen for useful discussions.  This work has been financially supported by the Research Foundation Flanders (FWO).





\end{document}